\documentclass[prb,reprint,preprintnumbers,amsmath,amssymb,showpacs,superscriptaddress,citeautoscript]{revtex4-1}
\usepackage{amssymb}

\usepackage{epsfig}
\usepackage{graphicx}% Needed to include png/jpg/pdf figure files
\usepackage{amsmath}
\usepackage{bm}
\usepackage{color, soul}
\usepackage[normalem]{ulem}
\usepackage{siunitx}

\usepackage[breaklinks]{hyperref}%hyperef package does internal and external linking (labels, urls)
\usepackage[all]{hypcap}%addition to hyperref, it ensures figure is correctly displayed (and not only the caption) when link is clicked
\hypersetup{
    plainpages=false,
    bookmarks=false,         % show bookmarks bar?
    unicode=false,          % non-Latin characters in AcrobatĂŻÂżÂ˝s bookmarks
    pdfborder={0 0 0}
    pdftoolbar=true,        % show AcrobatĂŻÂżÂ˝s toolbar?
    pdfmenubar=true,        % show AcrobatĂŻÂżÂ˝s menu?https://www.overleaf.com/13403665mnpkhvmzmbnjhttps://www.overleaf.com/13403665mnpkhvmzmbnj
    pdffitwindow=false,     % window fit to page when opened
    pdfstartview={FitH},    % fits the width of the page to the window
    pdftitle={BulkCrystalMagnetoOptics},    % title
    pdfauthor={Michal Baranowski},     % author
    pdfproducer={LNCMI-CNRS-UGA-UPS-INSA}, % producer of the document
    pdfkeywords={High} {Magnetic} {Fields}, % list of keywords
    pdfnewwindow=true,      % links in new window
    linktoc=section,
    colorlinks=true,       % false: boxed links; true: colored links
    linkcolor=blue,          % color of internal links
    citecolor=red,        % color of links to bibliography
    filecolor=magenta,      % color of file links
    urlcolor=blue           % color of external links
}

\begin{document}

\title{Non equilibrium anisotropic excitons in atomically thin ReS$_2$}

\author{J.\ M.\ Urban}\thanks{These authors contributed equally to the work}
\affiliation{Laboratoire National des Champs Magn\'etiques Intenses, UPR 3228, CNRS-UGA-UPS-INSA, Grenoble and Toulouse, France}
\author{M.\ Baranowski}\thanks{These authors contributed equally to the work}
\affiliation{Laboratoire National des Champs Magn\'etiques Intenses,
UPR 3228, CNRS-UGA-UPS-INSA, Grenoble and Toulouse, France}
\affiliation{Department of Experimental Physics, Faculty of
Fundamental Problems of Technology, Wroclaw University of Science and Technology, Wroclaw, Poland}
\author{A.\ Kuc}
\affiliation{Wilhelm-Ostwald Institute f\"{u}r Physikalische und Theoretische Chemie, Universit\"{a}t Leipzig, Linn\'{e} Str. 2, 04103 Leipzig, Germany}
\affiliation{Helmholtz-Zentrum Dresden-Rossendorf, Abteilung Ressourcen\"{o}kologie, Forschungsstelle Leipzig, Permoserstr. 15, 04318, Leipzig, Germany}
\author{{\L}.\ K{\l}opotowski}
\affiliation{Institute of Physics, Polish Academy of Sciences, Al.\ Lotnik{\'o}w 32/46, 02-668 Warsaw, Poland}
\author{A.\ Surrente}
\affiliation{Laboratoire National des Champs Magn\'etiques Intenses, UPR 3228, CNRS-UGA-UPS-INSA, Grenoble and Toulouse, France}
\author{Y.\ Ma}
\affiliation{Wilhelm-Ostwald Institute f\"{u}r Physikalische und Theoretische Chemie, Universit\"{a}t Leipzig, Linn\'{e} Str. 2, 04103 Leipzig, Germany}
\author{D.\ W{\l}odarczyk}
\affiliation{Institute of Physics, Polish Academy of Sciences, Al.\ Lotnik{\'o}w 32/46, 02-668 Warsaw, Poland}
\author{A.\ Suchocki}
\affiliation{Institute of Physics, Polish Academy of Sciences, Al.\ Lotnik{\'o}w 32/46, 02-668 Warsaw, Poland}
\author{D.\ Ovchinnikov}
\affiliation{Electrical Engineering Institute and Institute of Materials Science and Engineering, \'{E}cole Polytechnique F\'{e}d\'{e}rale de Lausanne, CH-1015 Lausanne, Switzerland}
\author{T.\ Heine}
\affiliation{Wilhelm-Ostwald Institute f\"{u}r Physikalische und Theoretische Chemie, Universit\"{a}t Leipzig, Linn\'{e} Str. 2, 04103 Leipzig, Germany}
\affiliation{Helmholtz-Zentrum Dresden-Rossendorf, Abteilung Ressourcen\"{o}kologie, Forschungsstelle Leipzig, Permoserstr. 15, 04318, Leipzig, Germany}
\affiliation{Theoretical Chemistry, TU Dresden, Mommsenstr. 13, 01062 Dresden}
\author{D. K.\ Maude}
\affiliation{Laboratoire National des Champs Magn\'etiques Intenses, UPR 3228, CNRS-UGA-UPS-INSA, Grenoble and Toulouse, France}
\author{A.\ Kis}
\affiliation{Electrical Engineering Institute and Institute of Materials Science and Engineering, \'{E}cole Polytechnique F\'{e}d\'{e}rale de Lausanne, CH-1015 Lausanne, Switzerland}
\author{P.\ Plochocka}\email{paulina.plochocka@lncmi.cnrs.fr}
\affiliation{Laboratoire National des Champs Magn\'etiques Intenses, UPR 3228, CNRS-UGA-UPS-INSA, Grenoble and Toulouse, France}

\date{\today}

\begin{abstract}
We present a systematic investigation of the electronic properties of bulk and few layer ReS$_2$ van der Waals crystals using
low temperature optical spectroscopy. Weak photoluminescence emission is observed from two non-degenerate band edge excitonic
transitions separated by $\sim$ 20 meV. The comparable emission intensity of both excitonic transitions is incompatible with a
fully thermalized (Boltzmann) distribution of excitons, indicating the hot nature of the emission. While DFT calculations predict bilayer ReS$_2$ to have a direct fundamental band gap, our optical data suggests that the fundamental gap is indirect in
all cases.

\end{abstract}

\maketitle

%\section{Introduction}

Emerging transition metal dichalcogenides (TMDs) such as MoS$_2$, MoSe$_2$, MoTe$_2$, WS$_2$ and WSe$_2$ are attracting great
attention due to their remarkable electronic properties. In particular, the energy and the character of the band gap can be
easily tuned by varying the number of atomic layers in the
crystal \cite{mak2010atomically,splendiani2010emerging,aslan_linearly_2016,qiao_polytypism_2016,arora_highly_2017}. The two
dimensional confinement and reduced dielectric screening in the single and few layer limit result in significantly enhanced
exciton and trion binding energies \cite{mak2013tightly, ross2013electrical, chernikov2014exciton, he2014tightly}, while the lack
of inversion symmetry in the monolayer leads to valley-selective optical selection rules
\cite{xiao2012coupled,mak2012control,zeng2012valley,xu2014spin}.

Recently, layered semiconductors with in-plane anisotropy such as black phosphorus \cite{xia2014rediscovering,
carvalho2016phosphorene}, and rhenium dichalcogenides (ReX$_2$, where X stands for Se or S atoms) \cite{hafeez2017rhenium} have joined the family of intensively investigated van der Waals crystals. The sizable in-plane crystal asymmetry results in
anisotropic optical \cite{ho_dichroic_2007,ho_absorption-edge_1998,friemelt_optical_1993,ho_optical_1997,
liang_optical_2009,wolverson2014raman,arora_highly_2017,cuiTransient,aslan_linearly_2016,cui_nonlinear_2017,chenet_-plane_2015}, and electrical \cite{ho_dichroic_2007,cuiTransient, ho_-plane_1999, lin2015, tiong1999electrical} properties, which can be employed
in field effect transistors \cite{yang2014layer, corbet2014field, corbet2016improved, ovchinnikov2016disorder}, polarization
sensitive photodetectors \cite{zhang2016tunable}, and new plasmonic devices \cite{low2014plasmons}. The major advantage of Re dichalcogenides over black phosphorus is their stability under ambient conditions \cite{favron2015photooxidation}, making them potentially interesting for applications.

\begin{figure}[t!]
\centering
\includegraphics[width=0.95\linewidth]{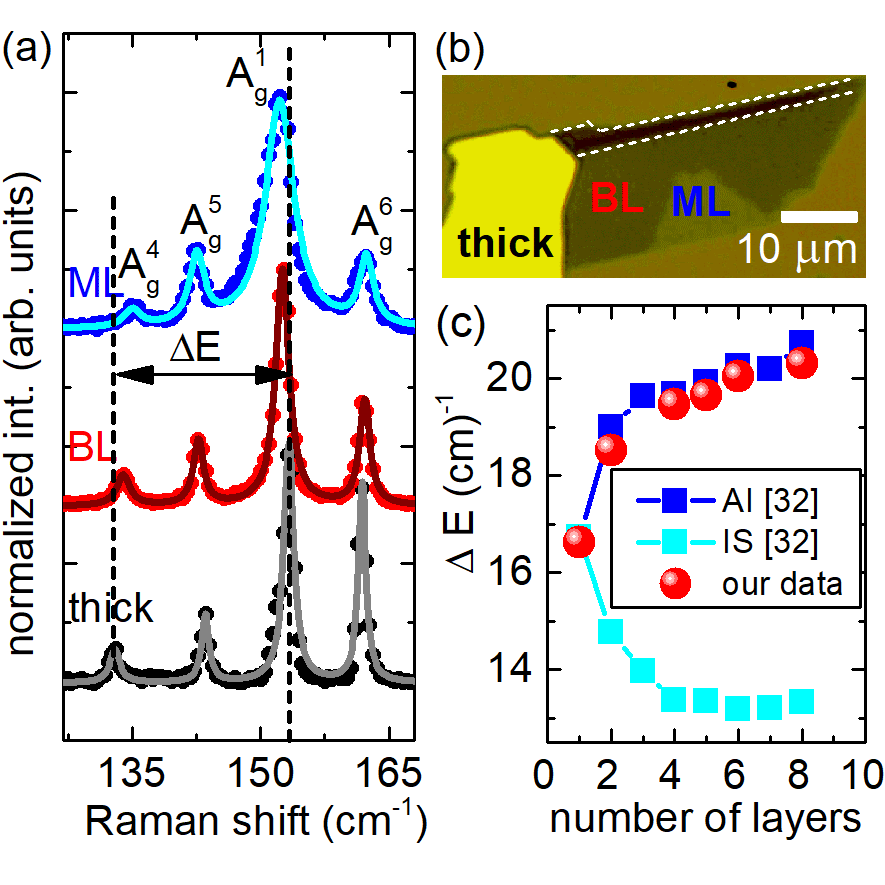}
\caption{(a) Representative Raman spectra from monolayer (blue), bilayer (red) and thick ($>8$ layers, bulk-like) (black) regions of the ReS$_2$ flake. The spectra have been normalized with respect to the A$^1_g$ intensity. We adopt the system used in previous work \cite{feng_raman_2015, mccreary_intricate_2017} to enumerate Raman modes. (b) Optical image of the investigated
ReS$_2$ flake. The white dashed line indicates a 4-6 layer thick area. (c) Difference $\Delta E=E_{\text{A}^1_\text{g}}-E_{\text{A}^4_\text{g}}$ between the A$^1_\text{g}$ and A$^4_\text{g}$ Raman modes as a function of the number of layers. Blue squares show data from  literature \cite{qiao_polytypism_2016}. Red spheres represent values from our measurements.} \label{fig:fig1}
\end{figure}

Unlike the more extensively studied Mo and W based dichalcogenides, Re based TMDs crystallize in the distorted 1T' structure (schematically shown in Fig.\ \ref{fig:fig1}(a)) of lower triclinic symmetry \cite{murray1994structure,kelty1994scanning,tongay2014monolayer, qiao_polytypism_2016}. In rhenium dichalcogenides, two non-degenerate direct excitons couple to light, as
observed in the reflectivity contrast and photoluminescence (PL) spectra \cite{huang_temperature_1997,aslan_linearly_2016,arora_highly_2017}. The strong linear polarization of the excitonic transitions provides a new degree of freedom to control the
optical response \cite{sim_selectively_2016,Sim2018} of this material. Despite the flurry of recent investigations of the ReS$_2$ and ReSe$_2$ \cite{hafeez2017rhenium}, knowledge about their fundamental electronic properties is extremely limited. For example, the nature of the fundamental band gap remains controversial.

Existing band structure calculations provide no consensus concerning the nature of the fundamental band gap \cite{tongay2014monolayer,arora_highly_2017,qiao_polytypism_2016,zhong_quasiparticle_2015, gehlmann_direct_2017, Webb2017}. Studies regarding the absorption
edge indicate that both materials have an indirect band gap in the bulk form
\cite{ho_absorption-edge_1998,ho_optical_1997,friemelt_optical_1993,liang_optical_2009, zelewski2017photoacoustic}. This assignment is supported by recent angle-resolved photoemission spectroscopy (ARPES) \cite{Webb2017} and photoemission electron
microscopy (PEEM) \cite{gehlmann_direct_2017} for the few- and monolayer crystals. However, the observed PL \cite{aslan_linearly_2016,arora_highly_2017,Gutierrez-Lezama2016} might suggest a direct nature of the band gap in this semiconductor. Intriguingly, a comparable emission intensity is observed from both excitonic states at cryogenic temperatures,
despite the fact that they are separated by a few tens of meV \cite{arora_highly_2017}. This clearly indicates a strong departure from a thermalized (Boltzmann) exciton distribution, which is expected if the direct excitons are the lowest states.

We have performed a systematic low temperature optical investigation of bulk and few layers ReS$_2$ combining Raman, PL and reflectivity contrast measurements. Our experimental results support the indirect character of the band gap from bulk down to single layer ReS$_2$ crystals. The PL has features of hot emission. Their observation, even
in the presence of an indirect band gap, is related to the relatively short radiative life time of direct excitons in ReS$_2$. Although the indirect nature of the gap is generally supported by density functional theory (DFT) band structure calculations, intriguingly, our DFT calculations predict a direct gap for bilayer ReS$_2$, while experimentally we detect only weak hot emission, characteristic for the indirect gap.

%\section{Methods}
Mono- and few-layer ReS$_2$ was obtained from bulk crystals (Hq-graphene) by scotch tape exfoliation onto a surface of
degenerately-doped Si covered with a 270\,nm thick layer of SiO$_2$. A flake containing monolayer, bilayer and a thick region ($>8$ layers) was selected for the investigation. The flake thickness was confirmed by optical contrast calibrated with Atomic Force Microscopy on flakes of different thicknesses (see Supporting Information). A microscope image of the flake is shown in Fig.\,\ref{fig:fig1}(b).

Raman measurements were performed in back scattering geometry under ambient conditions, at room temperature, and using a \SI{532}{\nano\meter} laser. The laser light was focused on a spot of approximately \SI{1}{\micro\meter} in diameter using a $100\times$ objective (numerical aperture, NA, 0.9). The spectral resolution of the setup was 0.5 cm$^{-1}$. The PL and reflectivity contrast measurements were performed in a standard $\mu$PL setup in
back-scattering geometry, with a long working distance $50\times$ objective having NA=0.55. The sample was excited by CW 532\,nm laser for PL or with white light from a tungsten-halogen lamp for reflectivity contrast measurements. The sample was mounted in a helium flow cryostat.

\begin{figure}[h!]
\centering
\includegraphics[width=0.9\linewidth]{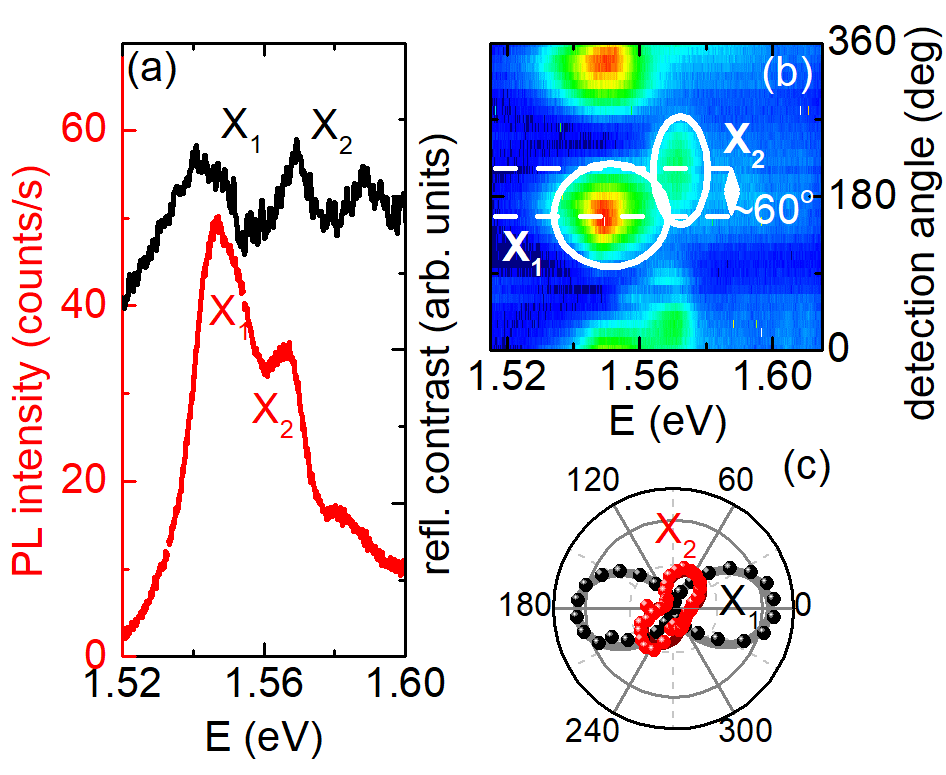}
\caption{(a) PL spectrum of bulk-like part of the sample measured at \SI{10}{\K} (red curve) together with the reflection contrast spectrum (black curve). (b) PL spectrum as a function of the linear polarization detection angle.  (c) Polar plot of the PL intensity of
X$_1$ (black balls) and X$_2$ (red balls) excitons as a function of polarization  detection angle together with the fitted curves.
For clarity, points on polar plot are shifted by about 25$^o$ comparing to results presented on (b) panel.}  \label{fig:fig2}
\end{figure}

%\begin{figure*}[t]
%\centering
%\includegraphics[width=0.99\linewidth]{Fig3.eps}
%\caption{Calculated DFT band structures of 1T' ReS$_2$ mono-, few-layer, and bulk. The arrows %indicate the optical (red) and the
%fundamental (black) electronic band gaps.}  \label{fig:fig3}
%\end{figure*}

Few-layer ReS$_2$ can be found in two different polytypes depending on the stacking order referred as isotropic or anisotropic \cite{qiao_polytypism_2016, hart_rhenium_2016}. They differ
in the relative orientation of the Re-Re bonds in neighboring layers. The number of layers and the stacking order can be determined using Raman spectroscopy \cite{qiao_polytypism_2016, mccreary_intricate_2017}. Fig.\,\ref{fig:fig1}(a) shows
typical Raman spectra measured for different thicknesses of the crystal. The colour coding of the spectra corresponds to that of the labels in the optical image in Fig.\,\ref{fig:fig1}(b). The pronounced A$_g^1$ mode at around 155\,cm$^{-1}$, related to the in-plane motion of Re and S atoms, together with lower intensity A$_g^4$, A$_g^5$ and A$_g^6$ are observed in all spectra. The four peaks were fitted with Lorentzian curves to precisely determine the characteristic vibration energy. With decreasing number of layers, the A$_g^1$ mode softens while the A$_g^4$ mode hardens. This is characteristic for anisotropic stacking order \cite{qiao_polytypism_2016}. Isotropic stacking order shows the opposite trend, \emph{i.e.}, the distance between
$A_g^1$ and $A_g^4$ mode decreases with increasing thickness. Therefore, the energy separation between $A_g^1$ and $A_g^4$  provides a reliable signature of the number of layers. By comparing our data with the previous work (see Fig.\ \ref{fig:fig1}(c)), we identify areas of monolayer and bilayer thicknesses, together with thicker regions at different positions on the flake.

A typical PL spectrum collected from a thick part of the crystal is presented in Fig.\,\ref{fig:fig2}(a). The two emission peaks at 1.546\,eV and 1.566\,eV  are non-degenerate direct excitonic transitions, previously observed in reflectivity contrast and PL \cite{aslan_linearly_2016, huang_temperature_1997, zelewski2017photoacoustic}. The free excitonic nature of these PL peaks is further supported by the reflectivity contrast, also plotted in Fig.\,\ref{fig:fig2}(a). In the reflectivity contrast,
both excitonic features are Stokes-shifted by $\simeq 10$\,meV with respect to the PL peaks. The full polarization dependence of the PL spectrum of the bulk-like sample is presented in Fig.\,\ref{fig:fig2}(b). The emission intensity of both excitons strongly depends on the detection polarization angle,  as demonstrated by the polar plot of Fig.\ \ref{fig:fig2}(c). These data are plotted in a non-polar plot in the Supplementary Information. We also show in the Supplementary Information the PL intensity of low energy exciton as a function of detection angle. These measurements demonstrate that the orientation of regions of different thicknesses within a same flake is the same. The variation of the PL intensity with the detection angle shows a characteristic figure of eight dependence
in the polar plot. This behavior is well described using $I_0+I_x\cos^2(\theta+\theta_0)$, from which we extract the angle between the maximum PL intensity for each exciton. In agreement with previous reports \cite{aslan_linearly_2016,sim_selectively_2016}, the  principal polarization axes of the excitons are rotated by $\simeq \SI{60}{\degree}$ with respect to each other.

\begin{figure*}[h]
\centering
\includegraphics[width=0.95\linewidth]{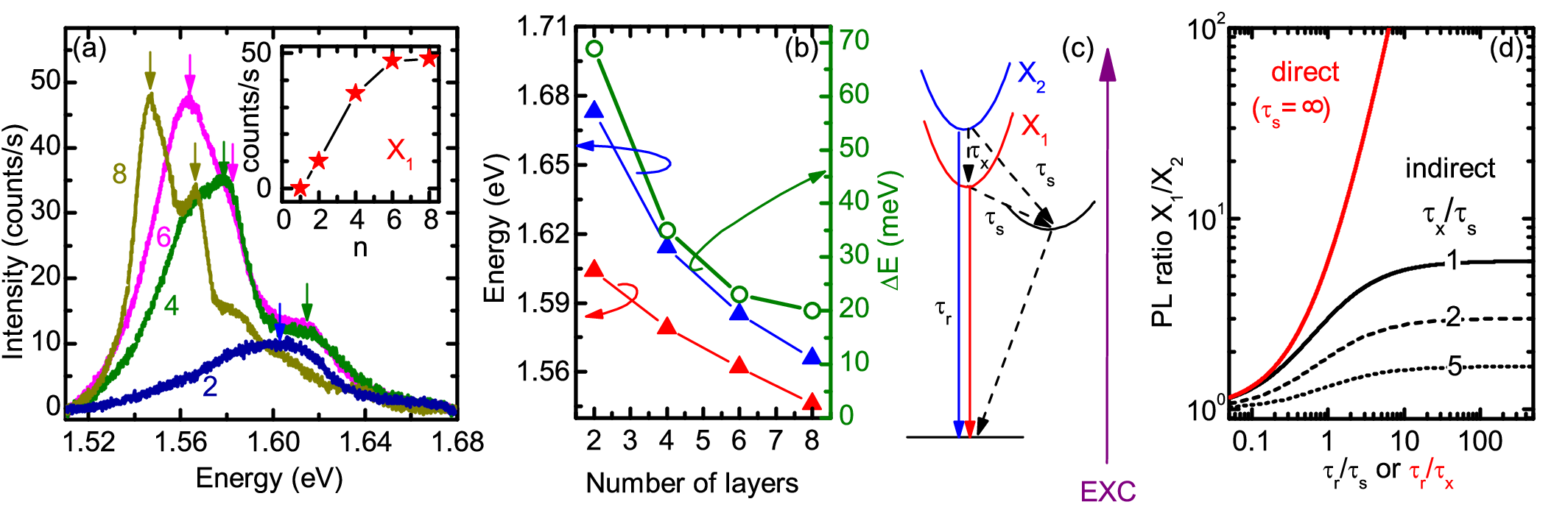}
\caption{(a) PL spectra measured at $T=10$\,K without polarization optics, for different number of layers. The arrows indicate the position of X$_1$ and
X$_2$ transitions. The inset shows the amplitude of the X$_1$ PL as a function of the number of layers $n$. (b) Dependence of
X$_1$ (red triangles) and X$_2$ (blue triangles) transition energy as a function of the number of layers and their energy separation.  (c) Scheme of carrier relaxation (dashed arrows) and radiative recombination process (red and blue arrows) in the presence of an indirect fundamental gap. (d)
Ratio of the X$_1$ to X$_2$ PL intensity calculated as function of $\tau_r/\tau_s$ (or $\tau_r/\tau_x$ for the direct band gap case with
$\tau_s = \infty$) by solving the rate equation model described in the Supplementary Information. For the indirect gap, the ratio was calculated for the three different values of
$\tau_x/\tau_s$ indicated.} \label{fig:fig3}
\end{figure*}

%To facilitate the interpretation of the optical data, we have performed DFT band structure calculations of the anisotropic ReS$_2$ crystal with different number of layers, as shown in Fig.\ \ref{fig:fig3}. The optical gap  at the $\Gamma$ point gradually increases with decreasing number of layers (see also Table 1 in Supporting Information). Intriguingly, the fundamental band gap is predicted to be direct only for the bilayer and becomes again indirect for the monolayer case. Our calculations are in agreement with bulk ReS$_2$ absorption measurements \cite{ho_absorption-edge_1998,ho_optical_1997,friemelt_optical_1993,liang_optical_2009, zelewski2017photoacoustic}, and recent photo-emission electron microscopy (PEEM) and GW calculations \cite{gehlmann_direct_2017}. However, the predicted direct nature of the the bilayer ReS$_2$ is not supported by the PL behavior, as we show in the next paragraphs.

Now we discuss how the evolution of the PL spectrum as a function of the number of layers, in the light of the theoretical predictions about the  nature of the bandgap. In the literature reports, one can find statements that the band gap nature is direct for all thicknesses\cite{tongay2014monolayer} as well as the prediction that the fundamental band gap is direct only for the bilayer\cite{gehlmann_direct_2017}. The latter is in agreement with our DFT calculations presented in Supplementary Information and absorption measurements for the bulk crystals\cite{ho_absorption-edge_1998,ho_optical_1997,friemelt_optical_1993,liang_optical_2009, zelewski2017photoacoustic}. However, none of these prediction is fully supported by the PL behavior as we show in the next paragraphs.

The evolution of the PL spectrum with the flake thickness is presented in Fig.\,\ref{fig:fig3}(a). As the number of layers
decreases, there is a systematic increase of the emission energy, related to the opening of the band gap at the $\Gamma$ point (see DFT calculation in Supplementary Information). The dependence of X$_1$ and X$_2$ transition energies versus number of layers is summarized in Fig.\,\ref{fig:fig3}(b).  In our sample, six and four layer thick ReS$_2$
corresponds to a narrow stripe (see Fig.\ \ref{fig:fig1}(b)). The PL spectrum measured in these regions receives  some contribution from the bilayer emission, visible around \SI{1.62}{\eV}. The twin-peak PL structure can be consistently observed in all places on our sample with
thickness $\geq 4$ layers (see Fig.\ \ref{fig:fig3}(a)). The PL from monolayer area of the flake was too weak to be observed, consistent with the predicted indirect nature of the gap. For bilayer, the PL spectrum is dominated by a weak, single  PL peak attributed to the X$_1$ exciton, with a weak, high energy shoulder corresponding to the X$_2$ emission. Given that the X$_2$ of the bilayer is weak in the unpolarized PL spectrum, we determine its transition energy based on the polarization resolved PL measurements shown in the Supplementary Information. The evolution of the amplitude of the X$_1$ PL peak with the number of layers is shown in the inset of Fig.\,\ref{fig:fig3}(a). The PL intensity from the bilayer clearly matches the trend, with no sign of enhanced emission, which is expected to be orders of magnitude stronger if the fundamental gap were direct (as observed in MoS$_2$\cite{mak2010atomically}).

The observation of both X$_1$ and X$_2$ emissions is at first sight surprising, since they are separated by at least 20\,meV, which is more than an order of magnitude larger than k$_{\text{B}}T$ at 10\,K ($\sim\SI{0.8}{\milli\eV}$). If the exciton population follows a Boltzmann
distribution, the intensity ratio between two peaks should be vanishingly small ($\sim 10^{-9}$), \emph{i.e.}, the higher energy X$_2$ emission should not be observed. The observed PL peaks must arise from hot PL involving emission from excitons that are not fully thermalized. For this to occur, the radiative lifetime must be sufficiently short, compared to the non-radiative
relaxation time. In addition, if there is an effective way to depopulate both X$_1$ and X$_2$, the intensity of both will be reduced, while the ratio of the intensities will be pushed towards unity. The presence of the indirect band gap below the direct band gap transitions provides such an effective depopulation path. The scheme of the carrier kinetics with the band structure and
relaxation paths are indicated schematically in Fig.\ \ref{fig:fig3}(c). Here, $\tau_r$ is the radiative lifetime of the direct excitons, $\tau_s$ is the lifetime for scattering to the indirect dark exciton state, and $\tau_x$ is the lifetime for scattering
from X$_2$ to X$_1$. The radiative recombination time of the direct exciton transition is not known for ReS$_2$. We expect, however, that it lies in the picoseconds time scale, as for other TMDs \cite{godde2016exciton,robert2016exciton,robert2017fine}. We estimate that
the PL intensity here is $\simeq 2$ orders of magnitude weaker than in direct band gap TMDs. Assuming $\tau_r \simeq 1$\,ps, this gives a reasonable ball park figure for the non radiative lifetime $\tau_s \simeq 10$\,fs. We have solved the rate equation model to see what conditions should be fulfilled to obtain the observed ratio of X$_1$ and X$_2$ PL intensity for the direct and indirect case, which is
plotted as a function of $\tau_r/\tau_s$ in Fig.\,\ref{fig:fig3}(d) (see Supplementary Information for more details). If the fundamental gap is direct, the ratio of the emission
intensities diverges, as expected for $\tau_r/\tau_s > 1$. When the fundamental gap is indirect, for $\tau_r/\tau_s
> 1$ the ratio reproduces nicely the experimental observations, saturating between $\simeq 1 - 6$ depending the value of $\tau_x/\tau_s$ used.  For the direct gap, reproducing the X$_1$ to X$_2$ intensity ratio observed experimentally would require
$\tau_r/\tau_x \leq 1$, which is not expected for semiconductors and incompatible with weak PL. We
therefore conclude that the observation of hot PL emission is a smoking gun signature of the indirect gap. Since both excitonic transitions are observed for all thicknesses, we deduce that in all cases the gap is indirect.

It is interesting to note that ab initio calculations at different levels of theory successfully predict the evolution of the band structure for TMDs, such as MoS$_2$ \cite{padilha2014nature, molina2013effect, ramasubramaniam2012large, qiu2013optical, mak2010atomically}.However, DFT and GW calculations do not seem to capture with sufficient accuracy the changes in the electronic structure of bilayer ReS$_2$. Both methods predict that the bilayer is a direct gap semiconductor\cite{gehlmann_direct_2017}. In MoS$_2$, the indirect band gap rapidly increases with the reduction of the layers number, because of significant out of plane component of the orbitals, which compose the indirect band edges \cite{padilha2014nature}. In the monolayer limit, this results in an indirect band gap that is much larger (hundreds of meV at the K points) than the direct one, because the latter is contributed mainly by orbitals confined in the transition metal plane, which are rather insensitive to the surroundings. In contrast, the dependence of the band structure of ReS$_2$ on the number of layers is not so pronounced, which is attributed to significantly weaker interlayer coupling  \cite{tongay2014monolayer}. Therefore, the difference between direct and indirect energy transition falls within the limit of calculation accuracy for the bilayer, which explains the inconsistency with experimental results. Additionally, the inclusion of Coulomb interaction might lead to an improved accuracy of the prediction of the nature of the fundamental band gap for all thicknesses.

%\section{Conclusion}

To conclude, we have presented a combined investigation of low temperature PL and reflectivity contrast of ReS$_2$ of different thicknesses with anisotropic stacking
order. The weak PL observed from two direct excitons (X$_1$ and X$_2$) separated by $\simeq20$\,meV indicates the hot nature of the PL, together with the presence of a lower lying indirect band gap. This conclusion is supported by the solution of a rate equation model. The weak intensity of the PL, observed for both excitons, is consistently observed down to bilayer thickness, which points to an indirect fundamental band gap of ReS$_2$, independent of the number of layers. For bilayer ReS$_2$, this
contradicts the predictions of our DFT calculations, as well as the GW calculations and PEEM investigation found in the literature \cite{gehlmann_direct_2017}. However, the difference between optical and fundamental gap is small, as the DFT and GLLB-SC calculations do not include the exciton binding energy.

\begin{acknowledgments}
This work was partially supported by BLAPHENE and STRABOT projects, which received funding from the IDEX Toulouse, Emergence
program,  ``Programme des Investissements d'Avenir'' under the program ANR-11-IDEX-0002-02, reference ANR-10-LABX-0037-NEXT, and
by the PAN--CNRS collaboration within the PICS 2016-2018 agreement. M.B. appreciates support from the Polish Ministry of Science
and Higher Education  within  the  Mobilnosc  Plus program (grant no. 1648/MOB/V/2017/0). AK and TH acknowledge funding of Deutsche Forschungsgemeinschaft via the FlagERA project HE 3543/27-1 and ZIH Dresden for providing computational resources.
\end{acknowledgments}

\bibliography{ReS2}

\end{document}